# Improved Algebraic Degeneracy Testing

**Jean Cardinal** ✉ 
Université libre de Bruxelles (ULB), Brussels, Belgium

**Micha Sharir** ✉ 
School of Computer Science, Tel Aviv University, Tel Aviv, Israel

## Abstract

In the classical linear degeneracy testing problem, we are given $n$ real numbers and a $k$-variate linear polynomial $F$, for some constant $k$, and have to determine whether there exist $k$ numbers $a_1, \ldots, a_k$ from the set such that $F(a_1, \ldots, a_k) = 0$. We consider a generalization of this problem in which $F$ is an arbitrary constant-degree polynomial, we are given $k$ sets of $n$ numbers, and have to determine whether there exist a $k$-tuple of numbers, one in each set, on which $F$ vanishes. We give the first improvement over the naïve $O^*(n^{k-1})$ algorithm for this problem (where the $O^*(\cdot)$ notation omits subpolynomial factors).

We show that the problem can be solved in time $O^*\left(n^{k-2+\frac{4}{k+2}}\right)$ for even $k$ and in time $O^*\left(n^{k-2+\frac{4k-8}{k^2-5}}\right)$ for odd $k$ in the real RAM model of computation. We also prove that for $k = 4$, the problem can be solved in time $O^*(n^{2.625})$ in the algebraic decision tree model, and for $k = 5$ it can be solved in time $O^*(n^{3.56})$ in the same model, both improving on the above uniform bounds.

All our results rely on an algebraic generalization of the standard meet-in-the-middle algorithm for $k$-SUM, powered by recent algorithmic advances in the polynomial method for semi-algebraic range searching. In fact, our main technical result is much more broadly applicable, as it provides a general tool for detecting incidences and other interactions between points and algebraic surfaces in any dimension. In particular, it yields an efficient algorithm for a general, algebraic version of Hopcroft's point-line incidence detection problem in any dimension.

**2012 ACM Subject Classification** Theory of computation → Design and analysis of algorithms; Theory of computation → Computational geometry

**Category** Algorithmic complexity

**Funding** *Micha Sharir*: Work by Micha Sharir was partially supported by ISF grant 260/18.

**Acknowledgements** This work was initiated during a visit of the authors at the Swiss Federal Institute of Technology (ETH) in Zürich, Switzerland.

# 1 Introduction

*Linear degeneracy testing* is a computational problem in which we are given $n$ real numbers and a $k$-variate linear polynomial $F$, for some fixed constant $k$, as input, and we seek $k$ numbers $a_1, \ldots, a_k$ from the input set such that $F(a_1, \ldots, a_k) = 0$ [4, 20]. An important special case is the 3SUM problem, in which $k = 3$ and $F$ is simply the sum of the three variables. This problem was first studied as a bottleneck problem in computational geometry [28], since it is reducible, in subquadratic time, to many problems in computational geometry, which are now known as *3SUM-hard problems*. It has acquired over the years the status of a basic hard problem in fine-grained complexity theory [45]. The case where $k$ is an arbitrary fixed constant and $F$ is the sum of the $k$ variables is a fixed-parameter version of the NP-complete subset sum problem, usually referred to as the $k$-SUM problem [12, 29, 33, 39]. Linear degeneracy testing has a higher-dimensional counterpart that is of crucial importance in computational geometry, called *affine degeneracy testing in* $\mathbb{R}^d$: Given a set of $n$ points in $\mathbb{R}^d$, decide whether there exist a $(d+1)$-tuple lying on a $(d-1)$-flat [21]. Many problems reduce to degeneracy testing in that sense, including degeneracy of Voronoi diagrams or checking for incidences between geometric objects. Quoting Ailon and Chazelle [4], "the list of problems (...) that can be reduced to degeneracy testing is nearly endless".

In this contribution, we consider an algebraic generalization of the scalar degeneracy testing problem that we call $k$-*POL*, in which the polynomial $F$ can be an arbitrary, bounded-degree $k$-variate polynomial. For simplicity, we consider the $k$-partite version of the problem, in which we are given $k$ sets of $n$ real numbers, and we consider $k$-tuples formed by picking one number in each set.[1]

▶ **Problem 1** ($k$-POL). *Given $k$ sets $A_1, \ldots, A_k$, each of $n$ real numbers, and a constant-degree real $k$-variate polynomial $F \in \mathbb{Z}[x_1, \ldots, x_k]$, determine whether $F$ vanishes on the Cartesian product $A_1 \times A_2 \times \cdots \times A_k$, that is, determine whether there exist $a_1 \in A_1, \ldots, a_k \in A_k$ such that $F(a_1, \ldots, a_k) = 0$.*

The $k$-POL problem can be solved in $O(n^{k-1} \log n)$ time using standard algebraic tools, such as the ones described in Basu, Pollack, and Roy [9]. To do so, we iterate over all $(k-1)$-tuples $(a_1, \ldots, a_{k-1})$ in $A_1 \times \cdots \times A_{k-1}$, find the $O(1)$ real roots of $F(a_1, \ldots, a_{k-1}, x) = 0$ for each such tuple, and sort the overall set of $O(n^{k-1})$ roots. Then, for each $a_k \in A_k$ we search $a_k$ in the resulting sequence, and declare a positive solution to the $k$-POL problem if and only if one of the searches succeeds. To this date, no better algorithm is known for this problem.

In this work, we present the first algorithm improving over this elementary method, that solves $k$-POL in time $O^*(n^{k-2+f(k)})$ for any $k > 2$, where $f(k) = 4/k + O(1/k^2)$ (a more precise expression for $f(k)$ is stated in the abstract and also given later), and the $O^*(\cdot)$ notation hides subpolynomial factors. We also show how to further improve on our bounds in the cases $k = 4$ and $k = 5$, in the nonuniform algebraic decision tree model. Before stating our results in full detail, we briefly review the previous works on both classical and more recent variants of the problem.

## 1.1 Previous related work

The best known upper bounds on the complexity of linear degeneracy testing in the uniform real RAM model are $O(n^{k/2} \log n)$ for even values of $k$, and $O(n^{\lceil k/2 \rceil})$ for odd values of $k$. The folklore algorithm yielding these upper bounds is referred to as the *meet-in-the-middle* algorithm, due to its similarity to meet-in-the-middle attacks in cryptography. In the nonuniform $k$-*linear* decision tree model, where only linear sign tests involving $k$ distinct input numbers are allowed and accounted for, it is known that $\Omega(n^{\lceil k/2 \rceil})$ queries are necessary [20]. Ailon and Chazelle [4] gave a similar lower bound for $t$-linear decision trees, where $t$ is slightly larger than $k$. Note that linear degeneracy testing has long been known to be solvable in polynomial time in the $n$-linear decision tree model (hence with unrestricted linear sign tests), with the degree of the polynomial in the bound independent of $n$ [37, 38]. The bound was subsequently improved by Cardinal, Iacono, and Ooms [12] and Ezra and Sharir [22]. In a remarkable breakthrough paper, Kane, Lovett, and Moran [31] finally managed to show that linear degeneracy testing could be solved in $O(n \log^2 n)$ time in the $2k$-linear decision tree model.

---

[1] The single set version of the problem can be recovered by letting all sets be identical. This, however, allows us to use the same number more than once. The reduction to the case where numbers can be picked at most once is nontrivial, see for instance [13, 17]. We will skip over these issues here.

The $k$-SUM problem is the special case of linear degeneracy testing in which the polynomial $F$ is simply the sum of the $k$ variables. The simplest instance of $k$-SUM is the well-known 3SUM problem, which is the case where $k = 3$. For a long time, the 3SUM problem has been conjectured to require $\Omega(n^2)$ time. It is only in 2014 that it was shown to be solvable in (slightly) subquadratic time by Grønlund and Pettie [30]. Further improvements were given by Chan [15]. Improvements on the decision tree complexity of 3SUM [30], and later in [27, 29], have been proposed before it was shown to be solvable with $O(n \log^2 n)$ 6-linear queries, as a special case of the aforementioned algorithm for linear degeneracy testing [31]. While some slightly subquadratic-time uniform algorithms exist, the existence of an algorithm solving 3SUM in time $O(n^{2-\delta})$ for some positive constant $\delta$ is still a major open problem. It has recently been shown that all nontrivial linear degeneracy testing problems for $k = 3$ are equivalent under subquadratic reductions [17]. This makes 3SUM one of the cornerstone computational problems in fine-grained complexity theory [45]. The 4-SUM problem is closely related to the so-called "Sorting $X + Y$" problem, in which we are asked to sort the pairwise sums of elements in two sets of $n$ numbers [25]. See also [33] for recent results on $k$-SUM.

The 3-POL problem, in which we look for three input numbers on which an arbitrary given bounded-degree polynomial $F$ vanishes, has first been studied in Barba et al. [8] in both the real RAM and the *algebraic decision tree* models. In an algebraic decision tree, we only count sign tests of constant-degree polynomials in the input data, and again forbid any other operation to access the data explicitly; see [10, 40] and below. As shown in [8], the 3-POL problem can be solved in this model with only $O^*(n^{12/7})$ sign tests, an improvement over the $O^*(n^2)$ uniform upper bound.

The 3-POL problem has also been studied for the case where the three input sets $A$, $B$, $C$ are sets of points in the plane. This is an interesting extension because it contains as a special case the problem of *collinearity testing*, the two-dimensional version of affine degeneracy testing, in which we want to determine whether $A \times B \times C$ contains a collinear triple. Collinearity testing is a classical 3SUM-hard problem in computational geometry [28] for which no subquadratic algorithm is known, even in the algebraic decision tree model; see [8, 7] for a discussion. In the uniform model, the problem can be solved in $O(n^2)$ time. The primitive operation needed to test for collinearity of a specific triple $(a, b, c)$ is the so-called *orientation test*, in which we test for the sign of the determinant

$$\begin{vmatrix} 1 & x_a & y_a \\ 1 & x_b & y_b \\ 1 & x_c & y_c \end{vmatrix},$$

which is a quadratic polynomial in the six coordinates of a triple of points in $A \times B \times C$. Consequently, it is natural, and in fact necessary, to use the more general algebraic decision tree model mentioned above. Partial results with subquadratic algorithms in the algebraic decision tree model, both for the general 3-POL problem in the plane and for collinearity testing, have been obtained by Aronov, Ezra, and Sharir [7].

For the more general problem of affine degeneracy testing in $\mathbb{R}^d$, Erickson and Seidel [21] proved a lower bound of $\Omega(n^d)$ on the number of *sidedness queries*, in which one asks whether a point lies above, on, or below some hyperplane. The lower bound is matched by well-known algorithms [19]. The existence of a better real RAM algorithm, that would use more sophisticated operations than only sidedness queries, is a major open problem.

| k        | 4       | 5   | 6   | 7         | 8   |
|----------|---------|-----|-----|-----------|-----|
| Exponent | 2.666...| 3.6 | 4.5 | 5.4545... | 6.4 |

■ **Table 1** Upper bounds on the complexity of $k$-POL for the first few small values of $k$.

## 1.2  Our results

We provide the first algorithm for the $k$-POL problem that achieves a polynomial-factor improvement over the naïve method.

▶ **Theorem 2.** *The $k$-POL problem is solvable in time $O^*\left(n^{k-2+\frac{4}{k+2}}\right)$ for even $k$ and $O^*\left(n^{k-2+\frac{4k-8}{k^2-5}}\right)$ for odd $k$ in the real RAM computation model.*

Note that the speedup factor over the simple $O(n^{k-1}\log n)$ algorithm mentioned earlier gets close to linear for large $k$. The exponents for small values of $k$ are given in Table 1. In the cases $k = 4$ and $k = 5$, we can be further improve those bounds, albeit only in the more powerful algebraic decision tree model.

▶ **Theorem 3.** *The $4$-POL problem is solvable in time $O^*(n^{21/8})$ in the algebraic decision tree computation model.*

▶ **Theorem 4.** *The $5$-POL problem is solvable in time $O^*(n^{210/59})$ in the algebraic decision tree computation model.*

The exponents in the nonuniform bounds in Theorems 3 and 4 are 2.625 and $\sim 3.56$, respectively, which improve on the corresponding uniform bounds or $k = 4$ (2.666...) and $k = 5$ (3.6) (see Table 1).

**Application to affine degeneracy testing in $d$-space.**

One motivation for studying the $k$-POL problem is the following restricted version of affine degeneracy testing in $\mathbb{R}^d$. Let $A_1,\ldots,A_k$ be $k$ sets of $n$ points in $\mathbb{R}^d$, where each of the sets $A_i$ lies on its own one-dimensional constant-degree algebraic curve $\gamma_i$. The goal is to decide, in the real RAM computation model, whether there exist $k$ points, one in each set, that lie on a common $(k-2)$-flat. We suppose that each $\gamma_i$ is polynomially parameterizable. That is, $\gamma_i$ is given by the equations $x_j = f_{i,j}(t)$, for $j = 1,\ldots,d$, $t \in \mathbb{R}$, where the $f_{i,j}$ are polynomials of some constant maximum degree. Hence, we may represent each set $A_i$ as a set of $n$ real numbers, which are the values of the parameter $t$ that define its points. An example of an instance with $n = 4$ and $k = 3$ is shown in Figure 1.

Up to a simple randomized preprocessing, we can assume that $k = d + 1$, that is, we reduce the problem to one in which we look for $d + 1$ points lying on a common $(d-1)$-flat. Indeed, take a generic projection $\pi : \mathbb{R}^d \to \mathbb{R}^{k-1}$ with respect to $S = A_1 \cup \cdots \cup A_k$, namely a projection that does not introduce additional degeneracies in $S$. Thus a $k$-tuple of points of $\pi(S)$ lie on a hyperplane in $\mathbb{R}^{k-1}$ if and only if its preimage in $S$ lie on a $(k-2)$-flat. Note that for natural choices of distributions, a *random projection* is generic with respect to $S$ with probability one (see for instance [6] for a discussion).

We now consider the reduced problem, and note that the condition that $d + 1$ points

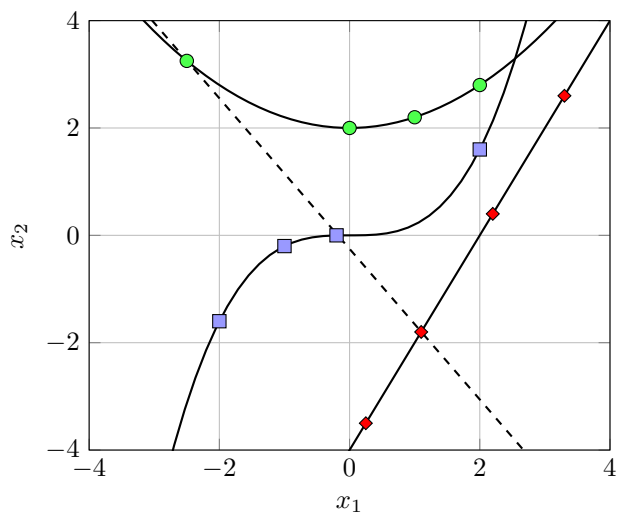

**Figure 1** An instance of the affine degeneracy testing in $\mathbb{R}^2$ for $n = 4$ and $k = 3$, where the three sets of points lie on three curves $\gamma_1, \gamma_2$ and $\gamma_3$ with $f_{1,1}(t) = f_{2,1}(t) = f_{3,1}(t) = t$, and respectively $f_{1,2}(t) = t^2/5 + 2$, $f_{2,2}(t) = t^3/5$, and $f_{3,2}(t) = 2t - 4$. The instance has a solution, indicated by the dashed line containing one point of each set.

$a_1 \in A_1, \ldots, a_{d+1} \in A_{d+1}$ lie on a common hyperplane is that the matrix

$$\begin{bmatrix} 1 & f_{1,1}(a_1) & f_{1,2}(a_1) & \cdots & f_{1,d}(a_1) \\ 1 & f_{2,1}(a_2) & f_{2,2}(a_2) & \cdots & f_{2,d}(a_2) \\ & & \cdots & & \\ 1 & f_{d+1,1}(a_{d+1}) & f_{d+1,2}(a_{d+1}) & \cdots & f_{d+1,d}(a_{d+1}) \end{bmatrix}$$

has determinant zero. This determinant is a bounded-degree polynomial in the input. Applying Theorem 2, we directly get the following result.

▶ **Corollary 5** (Constrained affine degeneracy testing in $\mathbb{R}^d$). *Let $A_1, \ldots, A_k$ be $k$ sets of $n$ points in $\mathbb{R}^d$, such that each of the sets $A_i$ lies on its own one-dimensional constant-degree algebraic curve, where all these curves are polynomially parameterizable. Then one can decide, in the real RAM computation model, whether there exists $k$ points, one in each set, that lie on a common $(k-2)$-flat, in time $O^*\left(n^{k-2+\frac{4}{k+2}}\right)$ for even $k$ and $O^*\left(n^{k-2+\frac{4k-8}{k^2-5}}\right)$ for odd $k$.*

We remark that a similar result can be obtained for more general constant-degree algebraic curves, using a more involved algebraic machinery, but we do not spell out the details of this generalization here. All the above algorithms rely on an algorithm for a general algebraic version of the incidence detection problem.

**Hopcroft's problem.**

Hopcroft's classical problem is that of determining, given two collections of $n$ points and of $n$ lines in the plane, whether some point lies on some line. An elegant algorithm relying on cuttings was proposed by Matoušek [34], with running time $n^{4/3}2^{O(\log^* n)}$. It has recently been slightly improved to $O(n^{4/3})$ by Chan and Zheng [14]. Our main technical result, given in Theorem 6, is an algebraic generalization of this result, in which we wish to detect incidences between points and algebraic surfaces of codimension 1 and of constant degree in arbitrary dimension. This involves a careful use of recent algorithmic methods for hierarchical polynomial partitions.

**Organization of the paper.**

The next section is dedicated to solving this generalized algebraic version of Hopcroft's problem. In Section 3, we apply this result to the $k$-POL problem and prove Theorem 2. In Section 4, we consider algorithms in the algebraic decision tree model and prove Theorems 3 and 4.

## 2 Hopcroft's Problem generalized

Our main technical result is of independent interest, and provides a broad generalization of the classical Hopcroft's problem [14, 34], originally formulated for points and lines in the plane.

### 2.1 Statement

▶ **Theorem 6** (General Hopcroft's problem). *Let $k = s + t$, for any suitable pair of positive integer constants $s$ and $t$, and let $F$ be a real $k$-variate polynomial of constant degree. For any set $P$ of $N$ points in $\mathbb{R}^t$, and any set $Q$ of $M$ points in $\mathbb{R}^s$, deciding whether there exists a pair $(p, q) \in P \times Q$ such that $F(p_1, \ldots, p_t, q_1, \ldots, q_s) = 0$ can be done in time*

$$O^*\left(M^{1-\frac{t-1}{ts-1}} N^{1-\frac{s-1}{ts-1}} + M + N\right)$$

*in the real RAM model. Furthermore, one can obtain within the same time bound a compact encoding of the signs $\mathrm{sign}(F(p_1, \ldots, p_t, q_1, \ldots, q_s)) \in \{0, +, -\}$ for all $(p, q) \in P \times Q$.*

Hopcroft's problem itself is a special case with $t = s = 2$ and $F(p_1, p_2, q_1, q_2) = p_1 q_1 + p_2 - q_2$ (with a suitable parameterization of the input lines). Our bound in this case is $O^*(M^{2/3} N^{2/3} + M + N)$, which is close to the best known upper bound $O(M^{2/3} N^{2/3} + M + N)$, due to Chan and Zheng [14].

The case $s = t = 6$ was stated by Aronov et al. [5] for another application. A special case of Theorem 6 with $t = s = 2$ and $N = M$ was also stated by Barba et al. [8] (Lemma 6.8). When $s \neq t$, we obtain an *asymmetric* incidence detection problem, with distinct primal and dual space dimensions. (For a simple instance of this setup, think of points and arbitrary circles in the plane.) See also Fox et al. [24] for related general incidence bounds.

### 2.2 Proof

To prove Theorem 6, we employ a modified version of the recent machinery of Matoušek and Patáková [36] for range searching with semi-algebraic sets in higher dimensions. The original study [36] gives, for $N$ input points in $\mathbb{R}^t$, a data structure of size $O^*(N)$ that can be constructed in $O^*(N)$ randomized expected time, so that a query with a constant-complexity semi-algebraic range can be answered in $O^*(N^{1-1/t})$ time. The main technical result on which their algorithm is based is the following lemma, which will be the main technical tool for our algorithm too. (We have changed some of the notations in the lemma statement to conform with the other notations used in this work.)

▶ **Lemma 7** (Matoušek and Patáková [36]). *For every integer $t > 1$ there is a constant $K$ such that the following hold. Given an $N$-point set $P \subset \mathbb{R}^t$ and a parameter $r > 1$, there are numbers $r_1, r_2, \ldots, r_t \in [r, r^K]$, positive integers $\ell_1, \ell_2, \ldots, \ell_t$, a partition*

$$P = P^* \cup \bigcup_{i=1}^{t} \bigcup_{j=1}^{\ell_i} P_{ij}$$

*of $P$ into disjoint subsets, and for every $i, j$, a connected set $S_{ij} \subseteq \mathbb{R}^t$ containing $P_{ij}$, such that $|P_{ij}| \leq n/r_i$ for all $i, j$, $|P^*| \leq r^K$, and the following holds:*

(⋆) *If $h$ is a $t$-variate real polynomial of bounded degree and $Z(h)$ is its zero set then, for every $i = 1, 2, \ldots, t$, the number of the sets $S_{ij}$ crossed by $Z(h)$ is at most $O(r_i^{1-1/t})$.*

*The implied constants depend on the maximum degree of the defining polynomials. Furthermore, the sets $P^*$, $P_{ij}$, and $S_{ij}$ can be constructed in time $O(nr^C)$ where $C = C(t)$ is a constant depending on $t$.*

**Proof of Theorem 6.** We map each point $q = (q_1, \ldots, q_s) \in Q$ to the surface

$$\sigma_q = \{(x_1, \ldots, x_t) \mid F(x_1, \ldots, x_t, q_1, \ldots, q_s) = 0\} \tag{1}$$

in $\mathbb{R}^t$. Let $\Sigma$ denote the resulting collection of these $M$ surfaces. Note that the surfaces of $\Sigma$ have $s$ degrees of freedom. If any of these surfaces is the entire $t$-space we terminate the algorithm right away with a positive outcome. Otherwise, the problem has been reduced to the problem of detecting an incidence between some point of $P$ and some surface of $\Sigma$.

Fix a sufficiently large constant parameter $r$. If $N \geq M$ we apply Lemma 7 to $P \subset \mathbb{R}^t$, and if $N \leq M$ we apply Lemma 7 to the set $Q$ of the points in $\mathbb{R}^s$ that represent the surfaces of $\Sigma$, where the points of $P$ are now mapped to dual surfaces in $\mathbb{R}^s$ in a manner symmetric to that in (1). Assume without loss of generality that $N \geq M$, and follow the notations in Lemma 7.

For each $i = 1, 2, \ldots, t$, each surface in $\Sigma$ crosses (intersects but does not contain) at most $O\left(r_i^{1-1/t}\right)$ sets $S_{ij}$. Denote by $\Sigma_{ij}$ the subset of surfaces that cross $S_{ij}$. If there is a surface $\sigma_q$ that fully contains $S_{ij}$ then we have an incidence between $\sigma_q$ and each point $p \in P_{ij}$ (assuming, as we may, that none of the sets $P_{ij}$ is empty). Otherwise, let $q_{ij} = |\Sigma_{ij}|$. Then from (⋆) we have, for each $i$,

$$\sum_{j=1}^{\ell_i} q_{ij} \leq cMr_i^{1-1/t},$$

for a suitable constant $c > 0$. For each $i = 1, \ldots, t$ and $j = 1, \ldots, \ell_i$, we face a subproblem involving $P_{ij}$ and $\Sigma_{ij}$, of respective sizes at most $N/r_i$ and $q_{ij}$, which we solve recursively, possibly switching to the dual setup, depending on which of these two sizes is larger. In addition, we have the leftover set $P^*$, which is of constant size, so we can detect an incidence between some point of $P^*$ and a surface of $\Sigma$ in $O(M)$ time. We stop the recursion when the size of one of the sets becomes smaller than some constant threshold $n_0$, and then solve the problem using brute force, in time linear in the size of the other set. Note that at each recursive step we can, in addition, keep track of the signs of the values of $F(p, q)$ for the points $p$ in $P_{ij}$ and the points $q$ defining surfaces in $\Sigma$ that do not intersect $S_{ij}$.

Let $T(N, M)$ denote the maximum running time of the procedure for sets $P$, $\Sigma$ of respective sizes at most $N$, $M$.

▷ Claim 8. For any fixed positive integer constants $s$ and $t$, and real $\varepsilon > 0$, there is a constant $A$ such that for any set $P$ of at most $N$ points in $\mathbb{R}^t$ and any set $\Sigma$ of at most $M$ algebraic surfaces in $\mathbb{R}^t$ with $s$ degrees of freedom, we have

$$T(N, M) \leq A\left(M^{1-\frac{t-1}{ts-1}+\varepsilon}N^{1-\frac{s-1}{ts-1}+\varepsilon} + M^{1+\varepsilon} + N^{1+\varepsilon}\right). \tag{2}$$

**Proof.** The proof is by induction on $N$ and $M$. The base case is $N \leq n_0$ or $M \leq n_0$. In either case we have $T(N, M) = O((N + M)n_0) = O(N + M)$, which is subsumed by the right-hand side of (2) if we choose $A$ to be sufficiently large. Consider then the case where, say, $N \geq M > n_0$, and assume that (2) holds for all smaller values $N' \leq N$, $M' < M$ or $N' < N$, $M' \leq M$. Apply Lemma 7 to the primal setup, where $P \subset \mathbb{R}^t$ is the set of points; we would apply it in the dual setup, in which $Q \subset \mathbb{R}^s$ is the set of points that represent the surfaces of $\Sigma$ in the complementary case where $M \geq N > n_0$. We use the notations in that lemma. Since $r$ is a constant, the nonrecursive cost of the procedure is at most $B(N + M)$, where $B$ is a constant that depends on $r$ and on the various other constant parameters (this also accounts for the processing of $P^*$). By induction hypothesis, for each $i$ and $j$, the cost of the recursive processing of $P_{ij}$ and $\Sigma_{ij}$ is at most

$$A\left(p_{ij}^{1-\frac{s-1}{ts-1}+\varepsilon}q_{ij}^{1-\frac{t-1}{ts-1}+\varepsilon} + p_{ij}^{1+\varepsilon} + q_{ij}^{1+\varepsilon}\right), \tag{3}$$

where $p_{ij} = |P_{ij}|$. Observe that we have $p_{ij} \leq N/r_i$ for each $j$, and the quantities $N_i := \sum_{j=1}^{\ell_i} p_{ij}$ satisfy $\sum_{i=1}^{t} N_i \leq N$ (since the decomposition in Lemma 7 is into disjoint subsets). Recall also that $\sum_{j=1}^{\ell_i} q_{ij} \leq cMr_i^{1-1/t}$.

We now sum the bounds in (3) over $j$ for each fixed $i$. We first note that

$$\sum_{j=1}^{\ell_i} p_{ij}^{1+\varepsilon} \leq (N/r_i)^\varepsilon \sum_{j=1}^{\ell_i} p_{ij} = (N/r_i)^\varepsilon N_i.$$

Using Hölder's inequality, the sum is upper bounded by

$$A\left(\sum_{j=1}^{\ell_i} p_{ij}^{1-\frac{s-1}{ts-1}+\varepsilon}q_{ij}^{1-\frac{t-1}{ts-1}+\varepsilon} + \sum_{j=1}^{\ell_i} p_{ij}^{1+\varepsilon} + \sum_{j=1}^{\ell_i} q_{ij}^{1+\varepsilon}\right)$$

$$\leq A\left((N/r_i)^{1-\frac{t+s-2}{ts-1}+2\varepsilon}\sum_{j=1}^{\ell_i} p_{ij}^{\frac{t-1}{ts-1}-\varepsilon}q_{ij}^{1-\frac{t-1}{ts-1}+\varepsilon} + (N/r_i)^\varepsilon N_i + (cMr_i^{1-1/t})^{1+\varepsilon}\right)$$

$$\leq A\left((N/r_i)^{1-\frac{t+s-2}{ts-1}+2\varepsilon}N_i^{\frac{t-1}{ts-1}-\varepsilon}(cMr_i^{1-1/t})^{1-\frac{t-1}{ts-1}+\varepsilon} + (N/r_i)^\varepsilon N_i + (cMr_i^{1-1/t})^{1+\varepsilon}\right)$$

$$= A\left(N^{1-\frac{t+s-2}{ts-1}+2\varepsilon}N_i^{\frac{t-1}{ts-1}-\varepsilon}M^{1-\frac{t-1}{ts-1}+\varepsilon} \cdot \frac{c^{1-\frac{t-1}{ts-1}+\varepsilon}}{r_i^{(1+\frac{1}{t})\varepsilon}} + (N/r_i)^\varepsilon N_i + (cr_i^{1-1/t})^{1+\varepsilon}M^{1+\varepsilon}\right)$$

$$\leq A\left(N^{1-\frac{s-1}{ts-1}+\varepsilon}M^{1-\frac{t-1}{ts-1}+\varepsilon} \cdot \frac{c^{1-\frac{t-1}{ts-1}+\varepsilon}}{r_i^{(1+\frac{1}{t})\varepsilon}} + (N/r_i)^\varepsilon N_i + (cr_i^{1-1/t})^{1+\varepsilon}M^{1+\varepsilon}\right),$$

where in the last inequality we used the fact that $N_i \leq N$ for each $i$.

We then sum these bounds over $i = 1, \ldots, t$, add the nonrecursive cost $B(M + N)$, and obtain the overall upper bound (recalling that $r_i \geq r$ for each $i$)

$$A\left(N^{1-\frac{s-1}{ts-1}+\varepsilon}M^{1-\frac{t-1}{ts-1}+\varepsilon} \cdot \frac{tc^{1-\frac{t-1}{ts-1}+\varepsilon}}{r^{(1+\frac{1}{t})\varepsilon}} + \frac{N^{1+\varepsilon}}{r^\varepsilon} + \sum_{i=1}^{t}(cr_i^{1-1/t})^{1+\varepsilon}M^{1+\varepsilon}\right) + B(M+N). \tag{4}$$

For $r$ chosen sufficiently large, the factors $\frac{tc^{1-\frac{t-1}{ts-1}+\varepsilon}}{r^{\left(1+\frac{1}{t}\right)\varepsilon}}$ and $\frac{1}{r^\varepsilon}$ are both smaller than $1/4$. The only problematic factor is $Z := \sum_{i=1}^{t}(cr_i^{1-1/t})^{1+\varepsilon}$, which in general will be larger than 1. To address this issue, one can easily verify that, for $M \leq N$, we have

$$M^{1+\varepsilon} \leq \frac{1}{N^{\frac{(s-1)(t-1)}{st-1}+\varepsilon}} \cdot N^{1-\frac{s-1}{ts-1}+\varepsilon} M^{1-\frac{t-1}{ts-1}+\varepsilon}.$$

That is,

$$ZM^{1+\varepsilon} \leq \frac{Z}{N^{\frac{(s-1)(t-1)}{st-1}+\varepsilon}} \cdot N^{1-\frac{s-1}{ts-1}+\varepsilon} M^{1-\frac{t-1}{ts-1}+\varepsilon},$$

and we can make the factor $\frac{Z}{N^{\frac{(s-1)(t-1)}{st-1}+\varepsilon}}$ smaller than $1/4$, assuming that $n_0$ is sufficiently large, an assumption that, as already noted, affects the choice of $A$.

Substituting these bounds into (4), we obtain

$$T(N,M) \leq A\left(\frac{1}{2}N^{1-\frac{s-1}{ts-1}+\varepsilon} M^{1-\frac{t-1}{ts-1}+\varepsilon} + \frac{1}{4}N^{1+\varepsilon}\right) + B(M+N),$$

which, by choosing $A$ sufficiently large, is smaller than the right-hand side of (2). This establishes the induction step and thereby completes the proof of the lemma for the case $M \leq N$. The complementary case $M \geq N$ is treated in a fully symmetric manner. This completes the proof of Claim 8. ◀

Switching back to the $O^*(\cdot)$ notation, we have established Theorem 6. ◀

## 3 An improved real RAM algorithm for algebraic degeneracy testing

We can now apply Theorem 6 to obtain an improved algebraic degeneracy testing algorithm. We first briefly summarize the best known algorithm for $k$-SUM.

### 3.1 The meet-in-the-middle algorithm for $k$-SUM

The meet-in-the-middle algorithm for $k$-SUM, assuming $k$ even, proceeds by computing the $n^{k/2}$ sums of all $(k/2)$-tuples of numbers in the first $k/2$ sets, as well as all $n^{k/2}$ sums of $(k/2)$-tuples from the last $k/2$ sets. It then searches a pair of opposite sums that sum to $0$, by sorting each of the collections of sums, in time $O(n^{k/2} \log n)$, and then by merging the sequence of the former sums with the negated sequence of the latter. When $k$ is odd, we sort the collection of sums of the $(k-1)/2$-tuples composed from numbers of the first $(k-1)/2$ sets, and of sums of the $(k-1)/2$-tuples composed from the last $(k-1)/2$ sets, in time $O(n^{(k-1)/2} \log n)$. Then, for each number $x \in A_{(k+1)/2}$, we shift the negated sequence of the latter sums by $-x$, then merge it with the former sequence to detect a coincident pair of values, which implies a positive answer. This takes time $O(n^{(k+1)/2}) = O(n^{\lceil k/2 \rceil})$ overall. This algorithm, for odd values of $k$, is hinted at by Erickson [20], and described by Ailon and Chazelle [4].

### 3.2 An algebraic meet-in-the-middle algorithm

Our algorithm can be seen as an algebraic generalization of this elementary method. We note that in the very special case where $k$ is even, say, and $F$ has the form $G(F_1(x_1,\ldots,x_{k/2}), F_2(x_{k/2+1},\ldots,x_k))$ for suitable constant-degree polynomials $F_1$, $F_2$ and $G$, the meet-in-the-middle algorithm can be straightforwardly generalized to solve the $k$-POL problem within the same time bound. In general, however, $F$ does not have this separation-of-variables property, and we have to resort to a more involved algorithm, with the running time asserted in Theorem 2.

**Proof of Theorem 2.** Let $A_1, \ldots, A_k$ and $F$ be as defined in the formulation of the $k$-POL problem, and let $t$, $s$ be integers satisfying $t + s = k$. Define $P = A_1 \times \cdots \times A_t$ as a set of $N = n^t$ points in $\mathbb{R}^t$, and $Q = A_{t+1} \times \cdots \times A_k$ as a set of $M = n^s$ points in $\mathbb{R}^s$. We now apply Theorem 6 with these values, and obtain a main term in the running time bound that is proportional to

$$n^{s - \frac{st-s}{st-1}} n^{t - \frac{st-t}{st-1}} = n^{k-2 + \frac{k-2}{st-1}},$$

up to some subpolynomial factors. It is easily checked that this term dominates the other terms $n^s$ and $n^t$ (for $1 \leq s, t \leq k-1$). The running time bound is therefore $O^*\left(n^{k-2+\frac{k-2}{st-1}}\right)$. To minimize this bound, $s$ and $t = k - s$ should be as close to $k/2$ as possible. Thus for even values of $k$, we take $t = s = k/2$ and obtain the bound

$$O^*\left(n^{k-2+\frac{4}{k+2}}\right),$$

which is indeed an improvement over the simpler $O^*(n^{k-1})$ solution discussed earlier. For odd values of $k$, we take $t = (k-1)/2$ and $s = (k+1)/2$, and obtain the bound

$$O^*\left(n^{k-2+\frac{4k-8}{k^2-5}}\right),$$

again an improvement. This proves Theorem 2. ◀

## 4 Improved algorithms in the algebraic decision tree model

We present faster algorithms in the algebraic decision tree model for the special cases of 4-POL and 5-POL, and prove theorems 3 and 4. In the algebraic decision tree model we only count the number of sign tests of constant-degree polynomials in the input data; all other operations are free of charge, but they are not allowed to explicitly access the real numbers in the input, and can only manipulate them via the results of the sign tests. We refer to [10, 41, 44] for seminal works on this model, and to [5, 8, 12, 31] for recent results under this model. Before giving a description of the algorithms, we briefly recall the point location data structure described in Aronov et al. [5], that will be used in both algorithms.

### 4.1 A simple point location data structure

Let $\Gamma$ be a finite collection of $x$-monotone constant-degree algebraic curves in the plane, and let $\mathcal{A}(\Gamma)$ denote the arrangement of those curves. We define the *order type* of $\mathcal{A}(\Gamma)$ as the following information:
1. The vertical order of the curves at $x = -\infty$.
2. For each curve $\gamma \in \Gamma$, the left-to-right order of the intersection points of $\gamma$ with the other curves of $\Gamma$, where each point is tagged by its *index*, which is the number of intersections of the same pair of curves that lie to its left.

We use the following lemma.

▶ **Lemma 9** (Aronov et al. [5]). *Let $\Gamma$ be a finite collection of $x$-monotone constant-degree algebraic curves in the plane. Using only the order type of $\mathcal{A}(\Gamma)$, one can construct a data structure that allows to answer point location queries in $\mathcal{A}(\Gamma)$ in time $O(\log^2 |\Gamma|)$.*

The data structure itself is the one proposed by Lee and Preparata [32] (see also [18]). It stores the information related to the levels of the arrangement, each consisting of a sequence of vertices defined as intersections of pairs of curves. Each such intersection point $q$ is encoded by a triple $(i, j, k)$, indicating that this is the $k$th intersection, in the left-to-right order, of the curves $\gamma_i, \gamma_j \in \Gamma$ (where $k - 1$ is the index of $q$, as defined above). To answer a query, we perform a binary search on the levels of the arrangement. At each step of this binary search, we need to know whether the query point lies above or below (or on) some level. For this, we use a secondary binary search on the $x$-coordinates of the vertices of the level, which we can obtain as a left-to-right sorted sequence from the discrete information in the order type. The overall cost is therefore $O(\log^2 |\Gamma|)$. We refer to [5] for a more detailed description.

The main point in using this simple method, even though it is less efficient than standard point location techniques in the uniform RAM model, is that in order to construct the data structure, we only need to know the order type of the arrangement, and that this order type can be encoded by a predicate that involves only *triples* of curves of $\Gamma$ (or by pairs for the vertical order of the curves at $x = -\infty$). The small arity of this predicate is a crucial factor in the improvement of the running time in the algebraic decision tree model, compared to the uniform real RAM model.

The assumption that the curves are $x$-monotone can be lifted, since every constant-degree algebraic curve in the plane can be decomposed into $O(1)$ $x$-monotone arcs, where the constant involved only depends on the degree of the curve. This extension of the algorithm to the case of (bounded) arcs requires some care but is not difficult. Using a segment tree over the $x$-projections of the arcs, we only need to pay an extra logarithmic factor in the point location mechanism. This turns out to be negligible in the following developments, and we will simply use the fact that point location queries are answered in time $\log^{O(1)} |\Gamma|$.

## 4.2 Algorithm for 4-POL

We consider the special case $k = 4$ of the $k$-POL problem, with four input sets $A, B, C, D$ of $n$ real numbers each.

**Proof of Theorem 3.** Recall that we need to locate the points of $P = A \times B$ in the arrangement of the curves in $\Gamma = \{\gamma_{c,d} \mid (c, d) \in C \times D\}$, where $\gamma_{c,d} = \{(x, y) \in \mathbb{R}^2 \mid F(x, y, c, d) = 0\}$. We can safely assume these are indeed one-dimensional curves and not the entire plane.

We want to preprocess the arrangement $\mathcal{A}(\Gamma)$ into a point location data structure. However, instead of computing a single data structure for the whole set, we partition each of $C$ and $D$ into $n/g$ blocks, each consisting of $g$ consecutive points (maybe less for the last block), for a suitable parameter $g \ll n$ that we will fix later. Denote the $C$-blocks as $C_1, \ldots, C_{\lceil n/g \rceil}$ and the $D$-blocks as $D_1, \ldots, D_{\lceil n/g \rceil}$. We construct the point location data structure described in Section 4.1 for each of the $(n^2/g^2)$ cells of the form $C_i \times D_j$, namely for the curves $\gamma_{c,d}$ for $(c, d) \in C_i \times D_j$. We then search these structures with each of the $O(n^2)$ pairs in $A \times B$, to detect whether any such pair, regarded as a point in $\mathbb{R}^2$, lies on any of the curves. For each pair $(a, b) \in A \times B$, define the dual curve $\gamma_{a,b}^* = \{(z, w) \in \mathbb{R}^2 \mid F(a, b, z, w) = 0\}$. Again we may assume that this is a one-dimensional curve and not the entire $zw$-plane. We observe that we need to search with a pair $(a, b) \in A \times B$ only in the point location data structures corresponding to pairs of blocks $C_i \times D_j$ for which $\gamma_{a,b}^*$ crosses the axis-parallel box in the $wz$-plane that defines $C_i \times D_j$. (See Figure 2 for an illustration.) Observe that there are only $O(n/g)$ such pairs, and the cost of one query in one of the blocks of size $g^2$ is $\log^{O(1)} g$, so the total cost of the $n^2$ searches is

$$O^*\left(\frac{n^3}{g}\right). \tag{5}$$

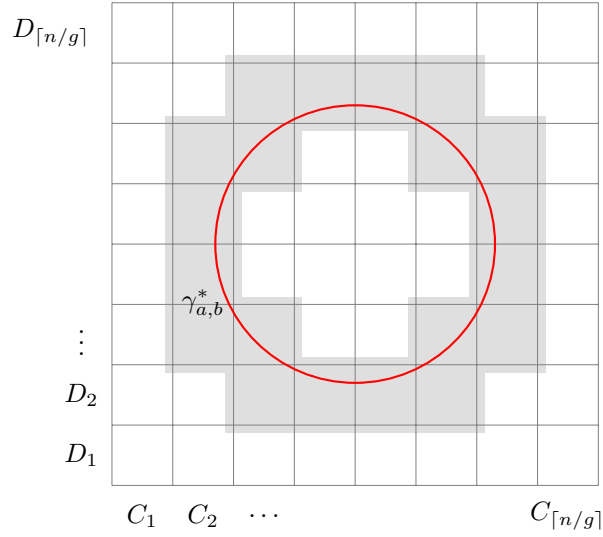

**Figure 2** Illustration of the algorithm for 4-POL. For each pair of values $(a,b) \in A \times B$, we test whether the dual curve $\gamma_{a,b}^*$ (here a circle of equation $w^2 + z^2 = 1$) is incident to any pair $(c,d) \in C \times D$. We only need to search the data structures for the pairs of blocks $C_i \times D_j$ whose boxes are intersected by $\gamma_{a,b}^*$, shown in gray. Note that the grid does not have to be uniform, but all boxes contain the same number $g^2$ of points of $C \times D$. The number of intersected boxes is $O(n/g)$.

It remains to describe the preprocessing phase, in which the point location data structures are constructed. We proceed as in the earlier recent works [5, 8], using the so-called Fredman's trick [30, 25, 26].

▷ **Claim 10.** The point location data structures for all the cells $C_i \times D_j$ can be constructed in time $O^*(n^{3/2}g^3)$ in the algebraic decision tree model.

**Proof of Claim 10.** In order to construct the point location data structure for the cell $C_i \times D_j$, we need to determine the order type of the arrangement of curves represented by points in $C_i \times D_j$. To determine this order type, we define a Boolean predicate $H_{k,k'}(c_1, c_2, c_3; d_1, d_2, d_3)$ with the following arguments:

1. a triple of curves $(\gamma_{c_1,d_1}, \gamma_{c_2,d_2}, \gamma_{c_3,d_3})$, where $c_1, c_2, c_3 \in C_i$ and $d_1, d_2, d_3 \in D_j$,
2. two positive integers $k, k'$ bounded by a constant depending on the degree of the curves.

$H_{k,k'}(c_1, c_2, c_3; d_1, d_2, d_3)$ determines the relative order along $\gamma_{c_1,d_1}$ of its $k$th leftmost intersection with $\gamma_{c_2,d_2}$ and its $k'$th leftmost intersection with $\gamma_{c_3,d_3}$. Specifically, $H_{k,k'}(c_1, c_2, c_3; d_1, d_2, d_3)$ is true if and only if the first intersection point lies to the left of the second point. (This suffices if we assume that no pair of intersection points coincide. Otherwise we add a predicate that is true when the two intersection points coincide.) Note that in practice $H_{k,k'}$ involves a number of quantifiers proportional to $k$ and $k'$, and eliminating these quantifiers is somewhat involved. Still, since the curves are of constant degree, all of this can be done in constant time using standard algebraic geometry techniques [9, 16].

To efficiently resolve all such comparisons, we split the predicate $H_{k,k'}(c_1, c_2, c_3; d_1, d_2, d_3)$ by considering $(c_1, c_2, c_3)$ as a point in $\mathbb{R}^3$ and $(d_1, d_2, d_3)$ as defining a constant-complexity semi-algebraic range

$$\sigma_{(d_1,d_2,d_3)} := \{(x_1, x_2, x_3) \mid H_{k,k'}(x_1, x_2, x_3; d_1, d_2, d_3) \text{ is true}\}.$$

For each of the $O(1)$ pairs $k, k'$, the number of points $(c_1, c_2, c_3)$ involved is equal to the number of blocks in $C$ multiplied by the number of ordered triples in each block, namely $(n/g) \cdot g^3 = ng^2$. Similarly, there are $ng^2$ ranges of the form $\sigma_{(d_1,d_2,d_3)}$. We now have a semi-algebraic batch range searching problem in $\mathbb{R}^3$, which has a symmetric dual version, in which the $c$-coordinates define ranges and the $d$-coordinates define points, also involving $ng^2$ points and $ng^2$ ranges. We can thus apply Theorem 6 with $N = M = ng^2$ and $t = s = 3$, since both points and curves have three degrees of freedom, and conclude that this problem can be solved in time

$$O^*\left((ng^2)^{3/2}\right) = O^*(n^{3/2}g^3).$$

This gives us the outcome of all the necessary comparisons to construct the point location structures, one for each pair $C_i, D_j$. The rest of the construction is free in the algebraic decision tree model. This proves Claim 10. ◀

It remains to (roughly) balance the cost in Claim 10 with that of the search phase given in (5); that is, ignoring subpolynomial factors, we set

$$\frac{n^3}{g} = n^{3/2}g^3 \quad \Rightarrow \quad g = n^{3/8}.$$

With this choice of $g$, the overall cost is $O^*(n^{21/8}) = O^*(n^{2.625})$, a polynomial improvement over the $O^*(n^{2.667})$ uniform algorithm. This proves Theorem 3. ◀

## 4.3 Algorithm for 5-POL

We now consider the case $k = 5$, with five input sets $A, B, C, D, E$ of $n$ real numbers each.

**Proof of Theorem 4.** We will locate the points of $P = A \times B$ in the arrangement of curves in $\Gamma = \{\gamma_{c,d,e} \mid (c, d, e) \in C \times D \times E\}$, where $\gamma_{c,d,e} = \{(x, y) \in \mathbb{R}^2 \mid F(x, y, c, d, e) = 0\}$. We first partition each of the three sets $C, D, E$ into blocks of $g$ consecutive values. We refer to the $i$th block of $C, D, E$ as $C_i, D_i, E_i$, respectively, where $i \in \{1, \ldots, \lceil n/g \rceil\}$.

Following the previous approach, we map each $(a, b) \in A \times B$ to the 2-surface $\sigma_{a,b}^* = \{(z, w, u) \in \mathbb{R}^3 \mid F(a, b, z, w, u) = 0\}$. One can show that $\sigma_{a,b}^*$ crosses only $O((n/g)^2)$ cells of the form $C_i \times D_j \times E_\ell$. (This property, and the corresponding property in the 4-POL algorithm, can be regarded as simple variants of the Schwartz–Zippel lemma; see [43, 46].) For each of the cells, we compute the point location data structure of Lemma 9, such that detecting an incidence can be performed in $\log^{O(1)} g$ time. Hence, the time spent on the query phase is

$$O^*\left(n^2 \cdot (n/g)^2\right) = O^*\left(\frac{n^4}{g^2}\right). \tag{6}$$

As for the preprocessing phase, we need to construct the point location data structure of Lemma 9 for each cell of the form $C_i \times D_j \times E_\ell$.

▷ **Claim 11.** The point location data structures for all the cells $C_i \times D_j \times E_\ell$ can be constructed in time $O^*(n^{42/17}g^{84/17})$ in the algebraic decision tree model.

**Proof of Claim 11.** We need to infer the order type of the arrangements in each of these cells. The order type can be inferred from the relative order of all pairs of intersections along a curve $\gamma_{c,d,e}$. This involves three curves $\gamma_{c_1,d_1,e_1}, \gamma_{c_2,d_2,e_2}, \gamma_{c_3,d_3,e_3}$, where $c_1, c_2, c_3 \in C_i$, $d_1, d_2, d_3 \in D_j$, and $e_1, e_2, e_3 \in E_\ell$ and amounts to determining the signs of a constant number of 9-variate real polynomials of the form $H_{k,k'}(c_1, c_2, c_3; d_1, d_2, d_3; e_1, e_2, e_3)$, defined in complete analogy to the predicates in Claim 10. This can be solved using again batch semi-algebraic range searching. We have $(n/g) \cdot g^3 = ng^2$ triples of the form $(c_1, c_2, c_3) \in \mathbb{R}^3$ and $(n/g)^2 \cdot g^6 = n^2 g^4$ 6-tuples of the form $(d_1, d_2, d_3; e_1, e_2, e_3) \in \mathbb{R}^6$. We can therefore apply Theorem 6 with $N = ng^2$, $M = n^2 g^4$, $t = 3$ and $s = 6$, and obtain the claimed running time of

$$O^*\left((n^2 g^4)^{15/17}(ng^2)^{12/17}\right) = O^*\left(n^{42/17} g^{84/17}\right).$$

◂

Balancing (roughly) the preprocessing cost of Claim 11 with the query cost in (6), we obtain

$$\frac{n^4}{g^2} = n^{42/17} g^{84/17} \quad \Rightarrow \quad g = n^{13/59},$$

yielding an overall complexity of $O^*\left(n^{4-26/59}\right) \simeq O^*\left(n^{3.56}\right)$, an improvement over the uniform bound $O(n^{3.6})$ obtained in the previous section. This proves Theorem 4. ◂

## Conclusion

We briefly mention some problems that we left open. An interesting question, for instance, is to give lower bounds for the $k$-POL problem that are asymptotically larger than that for the $k$-SUM problem. Note that $k$-SUM is known to not be solvable in time $n^{o(k)}$ under the exponential time hypothesis (ETH) [39]. Also, the techniques we use for algebraic decision trees do not seem to allow any speedup over the real RAM algorithm for $k \geq 6$. It would be interesting to design faster nonuniform algorithms for any value of $k$.